\documentclass[12pt]{article}
\usepackage{amsfonts}
\usepackage{latexsym}
\usepackage{showkeys}
\newtheorem{theorem}{Theorem}

\newlength{\dinwidth}
\newlength{\dinmargin}
\def\ben{\begin{displaymath}}
\def\een{\end{displaymath}}
\def\beq{\begin{equation}}
\def\eeq{\end{equation}}
\def\beqs{\begin{displaymath}}
\def\eeqs{\end{displaymath}}
\def\beqn{\begin{eqnarray}}
\def\eeqn{\end{eqnarray}}

\newcommand{\non}{\nonumber\\}

\newcommand{\iA}{{\scriptscriptstyle A}}
\newcommand{\iB}{{\scriptscriptstyle B}}
\newcommand{\iC}{{\scriptscriptstyle C}}
\newcommand{\iD}{{\scriptscriptstyle D}}

\newcommand{\iN}{{\scriptscriptstyle N}}
\newcommand{\iNminus}{{\scriptscriptstyle {N-1}}}
\newcommand{\izero}{{\scriptscriptstyle 0}}

\def\g{\gamma}
\def\l{\lambda}

\def\C{\mathbb {C}}
\def\Z{\mathbb {Z}}
\def\R{\mathbb {R}}
\def\N{\mathbb{N}}
\def\a{\alpha}
\def\d{\partial}
\def\gt{\tilde{\gamma}}
\def\gh{\hat{\g}}
\def\hsp{\hspace{0.5cm}}
\def\la{\label}
\def\f{\frac}

\def\L{{\cal L}}
\def\gm{\g_m}
\def\gn{\g_n}

\def\zi{{\cal Z}}

\def\sumablim{\sum^{K-1}_{\stackrel{\iA,\iB=0}{(\iA,\iB)\neq(0,0)}}}
\def\gm{\g_m}
\def\gn{\g_n}

\def\tr{\mathrm{tr}}
\def\tro{\stackrel{1}{\mathrm{tr}}}

\def\trt{\stackrel{2}{\mathrm{tr}}}
\def\Jmo{\stackrel1{J_m}}

\def\rot{\stackrel{12}{r}}
\def\rto{\stackrel{21}{r}}

\makeatletter
\@addtoreset{equation}{section}
\makeatother

\begin{document}
\title{Integrable systems related to elliptic branched coverings}
\maketitle
\begin{center}
{\large V.Shramchenko}

\vspace{1 cm}
Department of Mathematics and Statistics, Concordia University\\
7141 Sherbrooke West, Montreal H4B 1R6, Quebec, Canada
\end{center}
\vspace{1,5 cm}
\textbf{Abstract.}
The new integrable systems associated to the space of elliptic branched coverings are constructed. 
The relationship of these systems with elliptic Schlesinger's system \cite{Takasaki} is described. For the standard two-fold elliptic coverings the integrable system is written explicitly. The trigonometric degeneration of our construction is presented.

\newpage
\section{Introduction}
The most well-studied integrable systems like Korteweg-de Vries, non-linear Schlesinger, sin-Gordon \cite{ZMNP} appear as compatibility conditions of the auxiliary linear system
\beq
\Phi_x=U\Phi\;,\hspace{0.4cm}\Phi_y=V\Phi\;,
\label{linsystem}
\eeq
where $U\;,$ $V$ and $\Psi$ are matrix functions of $(x,y)$ and a constant (i.e., independent of $x$ and $y$) spectral parameter $\g\in\C\;.$ Matrices $U$ and $V$ for these systems are meromorphic functions of $\g$ with $(x,y)-$independent positions of poles. 

In 1978 Belinskii and Zakharov \cite{BelZak} and Maison \cite{Maison} discovered integrability of the Ernst equation
\begin{equation}
((x-y)G_x G^{-1})_{y} + ((x-y)G_{y} G^{-1})_{x}=0\;,
\la{Ernsteq}
\end{equation} 
where $G\in SU(1,1)/U(1)\;,$
which does not fit into this framework. Namely, the Ernst equation is a 
compatibility condition of the system (\ref{linsystem}) with matrices $U$ and $V$ of the form:
\beq
U=\frac{G_x G^{-1}}{1-\gamma}\;,\hspace{0.5cm}V=\frac{G_y G^{-1}}{1+\gamma}\;,
\label{UV}
\eeq
where the spectral parameter $\g$ is the function of $x\;,$ $y$ and a 
``hidden" (``constant") spectral parameter $\lambda\;:$
\beq
\g(\l,x,y)=\frac{2}{y-x}\Big(\frac{x+y}{2}-\lambda+\sqrt{(\l-x)(\l-y)}\Big)\;.
\eeq
Therefore the Ernst equation can be viewed as a ``deformation" of the principal chiral model (PCM) equations. For this model the matrices $U$ and $V$ have the same form (\ref{UV}) but $\g$ is a constant (independent of $(x,y)$) spectral parameter. 

The same equation (\ref{Ernsteq}) for $G\in SU(2)/U(1)$ plays the role of the Gauss-Weingarten system for the so-called Bianchi surfaces in $\R^3$ (surfaces of negative Gaussian curvature of special form \cite{Bianchi,Dima}). 

The general deformation scheme of linear systems of the type (\ref{linsystem}) was proposed in 1989 by Burtsev, Mikhailov and Zakharov \cite{BuMiZa}. Assuming that the spectral parameter $\g$ in (\ref{linsystem}) depends on $x$ and $y$, they derived a system of differential equations on $\g$ which provide a part of compatibility condition of the linear system (\ref{linsystem}). Solutions of the system for $\g$ was found in the recent work \cite{KokKor}; in this work $\g$ is given by the inverse map to the uniformization map of a rational (genus zero) $N-$fold branched covering of the Riemann sphere when the branch points of the covering are chosen to be  independent variables. In other words, a deformation of the linear system (\ref{linsystem}) was associated to the space of rational functions of degree $N$ with simple critical points. In the case of two-fold rational covering, if the matrix dimension equals $2$, this scheme leads to the Ernst equation. 

In \cite{KokKor} it was also shown how to generalize this approach to the Hurwitz spaces of genus $g\geq 2$ (spaces of meromorphic functions on the Riemann surface of genus $g$) for matrix systems. However, for the genus grater or equal to two it is difficult to present any explicit equations. The linear system  associated to a genus $g$ branched covering $\L$ has the following form \cite{KokKor}:
\beq
\frac{d\Psi}{d\l_m}=U_m\Psi\;,
\label{linsysKK}
\eeq
where the matrix $U_m(P,\{\l_m\})\;,$ $P\in\L$ has only one simple pole at the ramification point $P_m$ of the covering $\L$ and does not have any other singularities. Such a function exists on a genus zero surface, but for the higher genus it must be non-single valued. This means that 
for genus greater than one the matrices $U_m$ get some multiplicative and (or) additive transformations under tracing along topologically non-trivial cycles of the surface. These transformations depend on branch points of the covering, which makes the corresponding integrable system transcendently nonlinear. 

In genus one, however, it is possible to develop in detail a scheme analogous to the genus zero case and this is the purpose of the present paper. 

Consider the Hurwitz space $H_{1,N}\;,$ the space of $N-$fold genus one coverings of the Riemann sphere with simple ramification points (coverings consisting of $N$ copies of $\C P^1$ with $2N$ ramification points). Projections of the ramification points on the base of the covering are called the branch points; we assume them to be distinct and denote by $\l_1,\dots,\l_{2N}\;.$ Consider the Abel map $\nu:\L\to\C$  from the genus one covering $\L$ onto its fundamental domain in the complex $\g-$plane. We denote by $\g_1,\dots,\g_{2N}$ the images of the ramification points under this map. They satisfy the following equations  as functions of the branch points:
\beqn
\frac{\d\g_n}{\d\l_m}&=&-\a_m[\rho(\gn-\gm)+\rho(\g_m)]\;,\hskip0.7cm m\neq n\;,
\non\nonumber
\frac{\d\g_m}{\d\l_m}&=&\!\!\!\sum_{n=1,\, n\neq 
m}^{2N}\a_n [\rho(\g_m-\g_n)+\rho(\g_n)]\;,
\eeqn
where $\rho$ denotes the logarithmic derivative of the Jacobi theta function $\theta_1\;;$ $\a_m$ are some coefficients subject to the differential equations: 
\beqn
\frac{\d\a_n}{\d\l_m}&=&-2\,\a_n\a_m\rho'(\g_n-\g_m)\;,
\non\nonumber
\frac{\d\a_m}{\d\l_m}&=&\!\!\!\sum_{n=1,\, n\neq 
m}^{2N}2\,\a_n\a_m\rho'(\g_n-\g_m)\;.
\eeqn
On a covering of genus one the linear system (\ref{linsysKK}) can be written in terms of the elliptic $r-$matrix, whose transformations under tracing along non-trivial contours of the covering are given by similarity transformations independent of the branch points. Namely, in this paper we consider the linear system (\ref{linsysKK}) where matrices $U_m$ look as follows:
\beq
\stackrel1{U}_m(P)=\,\trt\Big(\rot(\nu(P)-\g_m)\stackrel2{J_m}\Big)
\label{linsystor}
\eeq
with some matrices $J_m(\{\l_k\})\;,$ $P\in\L\;.$ Here we consider all matrices as operators in the tensor product of two copies of $\C^K$: $\stackrel1{\!\!A}=A\otimes I$, $\;\stackrel2{\!\!A}=I\otimes A\;;$ the elliptic $r-$matrix $\stackrel{12}{r}$ is a linear operator in $\C^K\otimes\C^K$. The main result of this paper is the integrability of the following system:  
\beqn
\frac{\d\Jmo}{\d\l_n}&=&\!-\;\;\alpha_n\Jmo\rho^\prime(\g_m-\g_n)\;-\;\alpha_m\trt\Big(\stackrel{12}{r}\!^\prime(\gm-\gn)\stackrel2{J_n}\Big)\;
\non
&-&\Big[
\Jmo ,\trt\Big(\rot(\gm-\gn)\stackrel2{J_n}\Big)\Big]\;.
\label{jmsysintro}
\eeqn
It appears as compatibility condition of the linear system (\ref{linsysKK}), (\ref{linsystor}).
The systems (\ref{jmsysintro}) are a genus one analogs of the integrable systems constructed in \cite{KokKor}; they give elliptic generalizations of the Ernst equation (\ref{Ernsteq}).

We define the $\tau-$function for the integrable system (\ref{jmsysintro}) as follows:
\beq
\frac{\d\log\tau}{\d\l_m}=\,\frac{1}{2\alpha_m}\mathrm{tr}(J_m^2)\;.
\label{tauintro}
\eeq
This system is compatible as a corollary of (\ref{jmsysintro}). For the genus zero two-fold coverings this definition gives rise to one of the metric coefficients on the corresponding 
space-time \cite{KokKor}.

The non-linear integrable system (\ref{jmsysintro}), together with the associated linear system (\ref{linsysKK}), (\ref{linsystor}), turns out to be closely related with the elliptic Schlesinger system proposed by Takasaki \cite{Takasaki}. Namely, from each solution of the elliptic Schlesinger system we can obtain a solution of the system (\ref{jmsysintro}). For these solutions there is a simple link between $\tau-$function (\ref{tauintro}) and $\tau-$function of the elliptic Schlesinger system: 
\beq
\tau(\{\l_m\})=\,\prod_{j=\,1}^L\Big(\frac{\d\nu}{\d\l}(Q_j)\Big)^{\tr
A^2_j/2}\tau_{\mathrm Sch}\left(\{z_k\}\right){\big{|}_{z_k=\,\g(Q_k)}}\;,
\label{tautauintro}
\eeq
where $\{z_1,\dots,z_L\}$ is a set of points in the $\g-$plane which forms a part of monodromy data for the elliptic Schlesinger system; $Q_1,\dots,Q_L$ are points on the covering whose images under the Abel map $\nu$ are given by $z_1,\dots,z_L\;$ and whose projection on the $\l-$sphere do not depend on the branch points $\{\l_m\}\;;$ matrices $A_1,\dots,A_L$ solve the Schlesinger system; the variables $\tr A_j^2$ are integrals of the elliptic Schlesinger system. 

The paper is organized as follows. In section \ref{genuszero} we discuss the genus zero case and present a slight generalization of the scheme proposed in \cite{KokKor}. In section \ref{genusone} we derive auxiliary differential equations describing the dependence of the Abel map $\nu$ of the genus one covering on the branch points. Further, we introduce the linear system (\ref{linsysKK}), (\ref{linsystor}) and derive the integrable system (\ref{jmsysintro}) as its compatibility condition. Then we define the tau-function of the integrable system. Finally, we write explicitly the system (\ref{jmsysintro}) in the case of the simplest elliptic covering. Section \ref{Schles} is devoted to a description of the link of the integrable systems constructed in section \ref{genusone} with the elliptic Schlesinger system proposed by Takasaki \cite{Takasaki}. In section \ref{example} we describe the trigonometric degeneration of the constructed integrable systems (\ref{jmsysintro}). 

\section{Integrable systems related to space of rational functions.}
\label{genuszero}
The goal of this section is to describe integrable systems related to the space of rational functions.
We present a different version of the construction proposed in \cite{KokKor}. 
Consider the space of rational functions of degree $N$ with $2N-2$ critical points which have the following form:
\beq
R(\g)=\frac{a_\iN\g^\iN+a_{\iNminus}\g^\iNminus+\dots+a_\izero}{\g^\iN+b_{\iNminus}\g^\iNminus+\dots+b_\izero}\;.
\label{rational}
\eeq
The genus zero algebraic curve
\beqs
\l=R(\g)
\eeqs
can be realized as an $N-$fold branched covering $\L$ of the $\l-$sphere $\C P^1\;;$ a point $P$ of the covering is a pair $(\l,\g)\;.$  We denote by $\pi$ the projection operator from the covering onto the underlying $\l-$sphere: $\pi(P)=\l$. Functions (\ref{rational}) have $2N-2$ critical points counting multiplicities; according to the Riemann-Hurwitz formula, the genus of the corresponding covering $\L$ is zero. We assume the ramification points of the covering to be simple and finite; denote them by $P_1,\dots,P_{2N-2}\;.$
Their projections $\pi(P_m)=\l_m$ on the $\l-$sphere (the branch points) are critical values of the rational function $R(\g)$: $\l_m=R(\g_m)$, where $\{\g_m\}$ are critical points of the function $R\;,$ i.e., solutions of the equation $R^\prime(\g)=0$.  We assume all branch points $\l_1,\dots,\l_{2N-2}$ to be distinct. 

To each element $l$ of the fundamental group $\pi_1\left(\C\setminus\{\l_1,\dots,\l_{2N-2}\}\right)$ one can assign an element $\sigma_l$ of the symmetric group $S_N\;,$ which describes how the sheets of the covering permute when $\l$ goes along the contour $l\;.$ In this way we can assign to the covering $\L$ a representation of $\pi_1\left(\C\setminus\{\l_1,\dots,\l_{2N-2}\}\right)$ in $S_N\;.$ For the fixed number of sheets, type of branch points and the assigned representation, the covering is determined by positions of the branch points, i.e., $\{\l_m\}_{m=1}^{2N-2}$ gives a set of local coordinates on the Hurwitz space. Observe that this set has $2N-2$ elements whereas the corresponding rational function (\ref{rational}) is defined by $2N+1$ parameters. This is because any M\"{o}bius transformation in the $\g-$sphere (determined by three parameters),
\beq
\g\mapsto\frac{a\g+b}{c\g+d}\;,\hsp ad-bc=1\;,
\eeq
leaves positions of the branch points $\{\l_m\}$ invariant.

For our purposes we fix the coefficient $a_\iN$ in the nominator of the rational function to be a constant, say $a_\iN=1\;;$ then the rational function (\ref{rational}) becomes:
\beq
R(\g)=\frac{\g^\iN+a_{\iNminus}\g^\iNminus+\dots+a_\izero}{\g^\iN+b_{\iNminus}\g^\iNminus+\dots+b_\izero}\;;
\label{rationalnorm}
\eeq
and at infinity the following asymptotics takes place:
\beq
\l=1+\frac{\beta}{\g}+o(\frac{1}{\g})\;,\hsp {\mbox {as}}\;\;\;\g\sim\infty\;,
\label{infinity}
\eeq
where we denoted $\beta=a_\iNminus-b_\iNminus\;.$

We shall consider the critical points $\{\g_n\}$ of the rational function (\ref{rationalnorm}) as functions of its critical values $\{\l_n\}$. First, note that on the covering $\L$ there defined a one-to-one function $\nu:\L\to\C P^1$ such that $R(\nu(P))=\pi(P)\;;$ in particular, the images of ramification points are the critical points of the rational function: $\nu(P_m)=\g_m\;.$ The function $\nu(P)$ takes every value only once, thus $\nu(P)$ is holomorphic everywhere except the point which is mapped to infinity; then we can write the expansion of $\nu(P)$ with respect to the local parameter $\sqrt{\l-\l_m}$ in a neighbourhood of the ramification point $P_m$ (for any $m=1,\dots,2N-2$) as follows:
\beq
\nu(P)=\g_m+v_m\sqrt{\l-\l_m}+O(\l-\l_m)\;,\hsp P\to P_m\;.
\label{ratexp}
\eeq
Let us differentiate these expansions with respect to $\l_n$ and rewrite the result in terms of $\nu\;,$ using the relation $\sqrt{\l-\l_n}=(\nu-\g_n)/v_n+O((\nu-\g_n)^2)$ which follows from (\ref{ratexp}). We see that the function $\d\nu/\d\l_n$ is a meromorphic function of $\nu$ which has a first order pole at the point $\g_n$ and is regular at all other critical points, i.e.,
\beq
\frac{\d\nu}{\d\l_n}=\frac{\a_n}{\g_n-\nu}+f(\nu)\;,
\label{dertemp}
\eeq
where $\a_n=(v_n/2)^2\;,$ and $f(\nu)$ is a function regular everywhere except the point at infinity. We find the behavior of this function at infinity differentiating the asymptotics (\ref{infinity}) (which holds for $\g=\nu(P)$ since locally, in a neighbourhood of the preimage of infinity $P\sim\nu^{-1}(\infty)\;,$ the function $\g(\l)=\nu(P)$ gives the inverse to $R(\g)\;,\;\g\sim\infty$ ) with respect to $\l_n\;:$ 
\beq
0=\frac{\beta_{\l_n}}{\nu}-\nu_{\l_n}\frac{\beta}{\nu^2}+o(\frac{1}{\nu})\;,\hspace{0.4cm} {\mbox {as}}\;\;\;\nu\sim\infty\;.
\label{infder}
\eeq
This implies the following equations describing the dependence of the function $\nu$ on the critical values of the corresponding rational function (\ref{rationalnorm}):
\beq
\frac{\d\nu}{\d\l_n}=\frac{\a_n}{\g_n-\nu}+\frac{\beta_{\l_n}}{\beta}\nu+c_n\;,\hsp n=1,\dots,2N-2\;,
\label{gammadiff}
\eeq
with some functions $c_n=c_n(\{\l_k\})\;.$

The compatibility condition of the system (\ref{gammadiff}) gives the following system of differential equations for the critical points $\{\g_m\}$ of the rational function (\ref{rationalnorm}):
\beq
\frac{\d\g_m}{\d\l_n}=\frac{\a_n}{\g_n-\g_m}+\frac{\beta_{\l_n}}{\beta}\g_m+c_n\;,\hsp n\neq m\;.
\label{gammamdiff}
\eeq
\textbf {Remark.}
We get the same equations if instead of the rational function (\ref{rationalnorm}) consider the one of the form:
\beq
R(\g)=\beta\g+\delta+\sum_{k=1}^{N-1}\frac{a_k}{\g-b_k}\;,
\label{rational1}
\eeq
which can be obtained from (\ref{rationalnorm}) by a M\"obius transformation.
\vspace{0.3cm}

Consider now the following system of linear differential equations for a matrix-valued function $\Psi(P,\{\l_m\})\;\;(m=1,\dots,2N-2):$
\beq
\f{d\Psi}{d\l_m}(P)=\f{\g_0-\g_m}{\nu(P)-\g_m} G_{\l_m}G^{-1}\Psi(P)\;,
\label{Psisystem}
\eeq
where $\g_o=\nu(P_0)\;,$ the projection $\pi(P_0)=\l_0\in\C P^1$ of the point  $P_0$  is independent of all $\{\l_m\}\;;$ $G(\{\l_m\})$ is a matrix-valued function. 
The compatibility condition for (\ref{Psisystem}) is given by the following system of non-autonomous
(since all $\g_m$ and $\g_0$ are non-trivial algebraic functions of
$\{\l_m\}$) coupled PDE's:
\beq
\Big(\frac{\g_0-\g_m}{\beta} G_{\l_m} G^{-1}\Big)_{\l_n}=\Big(\frac{\g_0-\g_n}{\beta} G_{\l_n} G^{-1}\Big)_{\l_m}\;.
\la{hieint}
\eeq

The described construction of the integrable systems gives a realization of the scheme of Burtsev, Mikhailov, Zakharov \cite{BuMiZa} who derived the compatibility conditions for the deformed linear system of the type (\ref{linsystem}). They obtained differential equations on the variable spectral parameter of the linear system which form a part of the compatibility condition. 
It was shown in \cite{KokKor} that the function $\nu(P)$ is a solution of these differential equations. 

In the case of the two-fold coverings $(N=2)$ corresponding to the rational function of the form (\ref{rational1}) with $\beta=1\;,\;\delta=0$ (the normalization considered in \cite{KokKor})
system (\ref{hieint}) coincides with the Ernst equation (\ref{Ernsteq}) after the identification $\l_1=x$, $\l_2=y$ (see \cite{KokKor}).

There exists a well-known relationship between these rational two-fold coverings and the surface theory: the Gauss-Weingarten equation for a surface in $\R^3$ with the Gaussian curvature $K=-[\rho(x,y)]^{-2}$ can be written in the following form \cite{Dima}:
\beq
(\rho G_x G^{-1})_{y} + (\rho G_{y} G^{-1})_{x}=0\;,
\label{GWeq}
\eeq
for $G\in SU(2)/U(1)\;,$
which for the case of the Bianchi surfaces ($\rho(x,y)=x-y$) formally coincides with equation (\ref{Ernsteq}).

Here the natural question arises: are there other coverings for which the system (\ref{hieint}) takes the form of the Gauss-Weingarten equation for some surfaces? (Then it would be a new integrable case in surface theory.) This occurs if the system (\ref{hieint}) has 
the property $\g_0-\g_m=-(\g_0-\g_n)$ for some pair of indeces $m\;,n\;;$ that is
\beq
\frac{\g_m+\g_n}{2}=\g_0\;,
\label{gamma0}
\eeq 
where $\g_0=\nu(P_0)$ is the image of the point $P_0\in\L$ whose projection $\l_0$ on the $\l-$sphere does not depend on $\{\l_k\}\;.$ Existence of such systems is an open question. Since the covering is locally defined by $2N-2$ independent variables $\{\l_m\}\;,$ two additional parameters of the rational function (\ref{rationalnorm}) could be used to impose some relations on $\{\g_m\}\;.$
As it was already noted the freedom to choose these parameters corresponds to two M\"obius transformations in the $\g-$sphere: $\g\to a\g$ and $\g\to\g+b\;.$ 
But the condition (\ref{gamma0}) is invariant with respect to both of these transformations, which means that for the given degree $N$ of a rational function we do not have any freedom to impose  condition (\ref{gamma0}) for any pair of $m$ and $n\;.$ However, there is still a possibility that (\ref{gamma0}) holds for some rational coverings as in the case of $N=2\;.$

\section{Integrable systems related to elliptic branched coverings}
\label{genusone}
In this section we construct an elliptic analog of the integrable system (\ref{hieint}).

\subsection{Differential equations for images of ramification points of elliptic coverings in fundamental domain}

The Hurwitz space $H_{1,N}$ is the space of meromorphic functions of degree $N$ on Riemann surfaces of genus one. 
Consider a meromorphic double-periodic function $R$ of $\g\in\C$ with periods $1$ and $\mu\;$ and $N$ simple poles within the fundamental domain $T=\C /\{1,\mu\}\;.$ As a function on $T\;,$ $R(\g)$ has degree $N\;.$
The equation
\beq
\l=R(\g)
\label{ellcov}
\eeq
defines an $N$-fold branched covering  (we again call
it $\L$) of the Riemann sphere. A point $P$ of the covering is a pair: $P=(\l,\g)\;.$ According to the Riemann-Hurwitz formula, this covering has $2N$
ramification points counting multiplicities; we assume them to be simple and finite and denote by $P_1,\dots,P_{2N}$. Projections $\{\pi(P_m)\}$ of the ramification points onto the $\l-$sphere (the base of the covering) are called the branch points. They are given by critical values $\l_1,\dots,\l_{2N}$ of the meromorphic function $R(\g):$ $\l_m=R(\gh_m),$ where $\gh_1,\dots,\gh_{2N}$ are critical points of $R(\g),$ solutions of the equation $R^\prime(\g)=0\;.$ We assume the branch points to be distinct: $\l_m\neq\l_n$ for $m\neq n\;.$  
Our choice of the local parameters on $\L$ is standard: in a neighborhood of a ramification
point $P_m$ we take $x(P)=\sqrt{\l-\l_m}\;,\; P\in\L\;,\;P\sim P_m\;;$ in a neighborhood of a 
point at infinity on any sheet we take  $x=1/\l$; at any other point variable $\l$ itself is  used as a local coordinate.
To the covering $\L$ it is assigned a representation of the fundamental group $\pi_1\left(\C\setminus\{\l_1,\dots,\l_{2N}\}\right)$ in the following way. 
To each element $l$ of the fundamental group one can assign an element $\sigma_l$ of the symmetric group $S_N\;,$ which describes how the sheets permute when $\l$ goes along the contour $l$ on the base of the covering. We fix this representation of $\pi_1\left(\C\setminus\{\l_1,\dots,\l_{2N}\}\right)\;.$ Then for the fixed number of sheets and type of ramification points (we fix them to be simple) the covering is determined by positions of the branch points; thus we consider  $\{\l_m\}_{m=1}^{2N}$ as a set of local coordinates on the space of elliptic coverings.

Introduce on $\L$ some canonical basis of cycles $(a,b)\;.$ Denote by
${\bf v}(P)\;,\; P\in\L$ the holomorphic Abelian differential with normalized $a$-period:
\beq
\oint_{a}{\bf v} =1\;;
\eeq
for our covering it has the form:
\beq
{\bf v}(P)=\frac{d\l}{R^\prime(\g)}=d\g\;.
\label{elldiff}
\eeq
The integral over $b$-cycle gives the module $\mu$ of the elliptic Riemann
surface $\L$:
\beq
\mu=\oint_{b} {\bf v}\;.
\eeq
The function $\nu(P)$ which maps $\L$ onto the fundamental domain $T=\C/\{1,\mu\}$ is given by
the Abel map
\beq
\nu(P)=\,\int_{\infty^{(0)}}^P {\bf v}\;,
\label{abel}
\eeq
where we choose the initial point of integration to coincide with the
point at infinity on some (the ``zero''th)  sheet of the covering $\L\;$.
We denote the images of the ramification points under this map by $\g_m\;.$ They differ from the critical points $\{\gh_m\}$ of the function $R$ by a shift (corresponding to the choice of initial point of integration in (\ref{abel})) modulo the period lattice $\{k\mu+l\;;\;l\;,k\in \N\}\;:$
\beqn
\g_m\equiv\gh_m-c\;,&m=1,\dots,2N\;,
\label{crit}
\eeqn
where $c$ is the second coordinate of the point $\infty^{(0)}\in\L\;:$ $\;\infty^{(0)}=(\infty,c)\;.$

The Jacobi theta functions are given by
\beq
\theta[p,q](\g;\mu)=\sum_{m\in\Z}{\mathrm{exp}}\{\pi i\mu(m+p)^2+2\pi i\;(m+p)(\g+q)\}\;.
\label{thetadef}\eeq
We denote by $\rho(\g)$ the logarithmic derivative of theta-function $\theta_1(\gamma)=-\theta[1/2,1/2](\g)\;:$
\beq
\rho(\g)=\f{d}{d\g}\log\theta_1(\g)\;;
\eeq
it has the following periodicity properties:
\beq
\rho(\g+1)=\rho(\g)\;,\hskip1.0cm
\rho(\g+\mu)=\rho(\g)-2\pi i\;.
\label{roperiod}
\eeq
The derivative $\rho'(\g)$ coincides with the Weierstrass ${\cal P}$-function
up to a rescaling of the argument and an additive constant.

The following theorem describes the dependence of the map
$\nu(P)$ (\ref{abel}) on $\l$ and the branch
points $\{\l_m\}$; it provides an elliptic version of equations
(\ref{gammadiff}).
\begin{theorem} 
\label{thm1}
The function $\nu(\l,\{\l_m\})$ defined by (\ref{abel}) satisfies the following system of
differential equations:
\beqn
\f{\d\nu}{\d\l}&=&\,\sum_{k=\,1}^{2N} \a_k[\rho(\nu-\g_k)+\rho(\g_k)]\;,
\label{glint}\\
\f{\d\nu}{\d\l_m}&=&\,-\a_m[\rho(\nu-\g_m)+\rho(\g_m)]\;,\;\;\; m=1,\dots,2N\;,
\label{glmint}
\eeqn
where we denoted
\beq 
\a_m= \f{1}{2}v^2_m=\f{1}{2}\Big[\frac{{\bf v}(P)}{d\sqrt{\l-\l_m}}\Big|_{P=P_m}\Big]^2\;;
\la{amdef}
\eeq
and $\{\g_m\}$ are the images of the ramification points under the map $\nu\;:$ $\g_m=\nu(P_m)\;.$
\end{theorem}

\textbf{Remark.} The form (\ref{elldiff}) of the holomorphic normalized differential implies that $\alpha_m=[R^{\prime\prime}(\gh_m)]^{-1}\;.$
\vspace{0.3cm}

{\it Proof of theorem \ref{thm1}.} 
From (\ref{abel}) we see that the function $\nu(P)$ 
is holomorphic in a neighborhood of the ramification point $P_m\;$ and behaves as follows:
\beq
\nu(P)=\,\g_m+v_m\sqrt{\l-\l_m}+{\cal O}(\l-\l_m)\hspace{0.5 cm} 
{\rm as} \hspace{0.3cm}P\rightarrow P_m\;,
\label{ellexp}
\eeq
where  $\sqrt{\l-\l_m}$ is the local coordinate in a neighborhood of $P_m$, and $v_m$ is defined by (\ref{amdef}).
Therefore, in this neighborhood
\beqn
\frac{\d\nu}{\d\l}(P)&=&\,\frac{v_m}{2\sqrt{\l-\l_m}}+{\cal O}(1)\;,\\
\label{tempder}
\f{\d\nu}{\d\l_m}(P)&=&\,-\frac{v_m}{2\sqrt{\l-\l_m}}+{\cal O}(1)\;,\\
\label{tempder1}
\f{\d\nu}{\d\l_n}(P)&=&{\cal O}(1)\;,\hskip0.7cm n\neq m\;.
\eeqn
We rewrite these 
expansions in terms of the coordinate $\nu$ taking into account definition (\ref{amdef}) of $\a_m$ 
and the correspondence between local parameters $\nu-\g_m$ and $\sqrt{\l-\l_m}$ given by  (\ref{ellexp}):
\beq
\f{\d\nu}{\d\l}(P)=\,\frac{\a_m}{\nu-\g_m}+{\cal O}(1)\;,
\hspace{0,4cm}\f{\d\nu}{\d\l_n}(P)=\,-\delta_{mn}\frac{\a_m}{\nu-\g_m}+{\cal O}(1)\;,
\label{behavior}
\eeq
as $P\to P_m$.

The function $\nu(P)$ transforms as follows under
the tracing along basic cycles on $\L$: 
\beq
\nu(P^a)=\,\nu(P)+1\;,\hskip0.7cm
\nu(P^b)=\,\nu(P)+\mu\;, 
\label{perg}
\eeq
where  $\nu(P^a),\; \nu(P^b)$ denote the analytic continuation of $\nu(P)$
along $a-$ and $b-$cycles respectively. Therefore, the derivative $\nu_{\l}$ is periodic with respect to tracing along the basic cycles. Then the function $\nu_{\l}$ has periods $1$ and $\mu$ in the $\g-$plane. Its local behavior at the points $\g_m\;,\;m=1,\dots,2N$ is given by (\ref{behavior}). Hence, we conclude that $\sum^{2N}_{k=1}\a_k=0$ as sum of residues, and the derivative $\nu_{\l}$ can be
expressed as follows in terms of function $\rho\;$:
\beq
\f{\d\nu}{\d\l}(\nu)=\,\sum_{k=\,1}^{2N} \a_k\rho(\nu-\g_k)+const\;.
\label{temp*}
\eeq
To determine the constant in (\ref{temp*}) consider a neighborhood of $P=\,\infty^{(0)}\;.$ The Abel map (\ref{abel}) is zero at this point, $\nu(\infty^{(0)})=\,0\;,$ and we can write its behavior there as follows:
\beq
\nu(\l)=\,\frac{\alpha}{\l}+{\cal O}\Big(\frac{1}{\l^2}\Big)\hspace{0.4cm}
{\rm as}\hspace{0.4cm} \l\rightarrow\infty\;.
\label{atinfty}
\eeq
(Note that $\alpha\neq 0$ since we assume $\infty^{(0)}$ not to be a ramification point.) Therefore, for the $\l-$derivative we have in a neighbourhood of $P=\infty^{(0)}\;:$
\beqs
\f{\d\nu}{\d\l}(\l)=\,-\frac{\alpha}{\l^2}+{\cal O}\Big(\frac{1}{\l^3}\Big)\;,\hspace{0.4cm} \l\rightarrow\infty\;.
\eeqs
Rewriting as before this expansion in terms of the coordinate $\nu$ ((\ref{atinfty}) implies $\l\sim\a/\nu\;$) we see that $\nu_{\l}(\nu=\,0)=\,0\;:$
\beqn
\f{\d\nu}{\d\l}(\nu)=\,-\frac{\nu^2}{\alpha}+{\cal O}\left(\nu^3\right)\;,&{\mbox {as }}\;\;\nu(P)\to 0\;.
\nonumber
\eeqn
Therefore, (\ref{temp*}) turns into (\ref{glint}).

Consider now $\nu_{\l_m}$. In the $\g-$plane it has only one simple pole at $\nu=\g_m$ as it follows from (\ref{behavior}). The periodicity properties (\ref{perg}) of the Abel map imply that
\beq
\frac{\d\nu}{\d\l_m}(\nu+1)=\,\frac{\d\nu}{\d\l_m}(\nu)\hsp \mbox{and}\hsp 
\frac{\d\nu}{\d\l_m}(\nu+\mu)=\,\frac{\d\nu}{\d\l_m}(\nu)+\frac{\d\mu}{\d\l_m}\;.
\label{period}
\eeq
The function 
$-\a_m \rho(\nu-\g_m)+const\;$ satisfies the periodicity condition (\ref{period}) since, due to the Rauch variational formulas \cite{Rauch}, we have:
\beq
\frac{\d\mu}{\d\l_m}=\,\pi i v^2_m = 2\pi i \a_m\;.
\label{mupolambdam}
\eeq
To find the constant term we again put  $\nu=0\;,$ i.e.,   
$P=\,\infty^{(0)}$. Then  from the asymptotics 
(\ref{atinfty}) we see that $\nu_{\l_m}(\nu=\,0)=\,0$, which leads to (\ref{glmint}).

$\Box$

\textbf {Remark.} Equations (\ref{glmint}) can also be deduced from the
Rauch variational formulas for the differential {\bf v} \cite{Rauch}. 

Compatibility conditions of the system (\ref{glint}), (\ref{glmint})
imply the system of differential equations describing the dependence of $\{\g_m\}$ on the branch points $\{\l_m\}$ (the indeces run through the set $\{1,\dots,2N\}$): 
\beqn
\frac{\d\g_n}{\d\l_m}&=&-\a_m[\rho(\gn-\gm)+\rho(\g_m)]\;,\hskip0.7cm m\neq n\;,
\label{gnpolm}\\
\frac{\d\g_m}{\d\l_m}&=&\!\!\!\sum_{k=1,\, k\neq 
m}^{2N}\a_k [\rho(\g_m-\g_k)+\rho(\g_k)]\;.
\label{gmpolm}
\eeqn
The equations for residues $\a_m$ which also follow from the compatibility
of (\ref{glint}) and  (\ref{glmint}) look as follows:
\beqn
\frac{\d\a_n}{\d\l_m}&=&-2\,\a_n\a_m\rho'(\g_n-\g_m)\;;
\label{difnpolambdam}\\
\frac{\d\a_m}{\d\l_m}&=&\!\!\!\sum_{k=1,\, k\neq 
m}^{2N}2\,\a_k\a_m\rho'(\g_k-\g_m)\;.
\label{difmpolambdam}
\eeqn
In fact, equations (\ref{difnpolambdam}) and (\ref{difmpolambdam}) are
nothing but the Rauch variational formulas \cite{Rauch} for the
holomorphic differential ${\bf v}$.

\subsection{Integrable systems}
Denote the matrix dimension of our system by $K$.
The classical elliptic $r$-matrix is the following linear operator in the
tensor product of two copies of $\C^K$:
\beq
\stackrel{12}{r}(\g)=\sumablim w_{\iA\iB}(\g)\stackrel{1}{\sigma}\!_{\iA\iB}\stackrel{2}{\sigma}\!^{\iA\iB}\;,
\la{rmat}
\eeq
where  $w_{\iA\iB}$ 
are given by the combinations of Jacobi's theta functions (\ref{thetadef}) ($(A,B)\neq(0,0)$):
\beq
w_{\iA\iB}(\g)=\,\frac{\theta_{[\iA\iB]}(\g)\theta^\prime_{[00]}(0)}{\theta_{[\iA\iB]
}(0)\theta_{[00]}(\g)}\;;
\label{wAB}
\eeq
where we denote
\beqs
\theta_{[\iA\iB]}(\g)=\theta_{[\iA\iB]}(\g;\mu)=\theta\left[{\scriptstyle\frac{A}{K}-\frac{1}{2},\frac{1}{2}-\frac{B}{K}}\right](\g;\mu)\;.
\eeqs
All the $w_{\iA\iB}$ have a simple pole with unit residue at $\g=0$ and the following twist properties:
\beqn
w_{\iA\iB}(\g+1)=\epsilon^\iA w_{\iA\iB}(\g)\;,&w_{\iA\iB}(\g+\mu)=\epsilon^\iB w_{\iA\iB}(\g)\;,
\label{wtwists}
\eeqn
where $\epsilon=e^{2\pi i/K}\;.$
The matrices $\sigma_{\iA\iB}$ are the higher rank analogs of the Pauli matrices;
 they form a basis of $sl(K,\C)$ and are defined as follows (for ${(A,B)\neq(0,0)}$):
\beq
\sigma_{\iA\iB}=H^\iA F^\iB\;,
\label{sigmab}
\eeq
where $F$ is the diagonal matrix
\beqs
F={\mathrm{diag}}\{1,\epsilon,\epsilon^2,\dots,\epsilon^{K-1}\}\;,
\eeqs
and $H$ is the permutation matrix
\beqs
H=\left(\begin{array}{cccccc}
0&1&0&.&.&0\\
0&0&1&.&.&0\\
.&.&.&.&.&.\\
0&0&0&.&.&1\\
1&0&.&.&.&0\\ \end{array}\right).
\eeqs
These matrices satisfy the relations $\epsilon FH=HF$, and $F^K=H^K=I$.

Together with $\sigma_{\iA\iB}$ we introduce the dual basis $\sigma^{\iA\iB}$:
\beq
\sigma^{\iA\iB}=\frac{\epsilon^{-\iA\iB}}{K}\sigma_{-\iA,-\iB}\;,
\eeq
such that
\beq
\tr\left(\sigma_{\iA\iB}\sigma^{\iC\iD}\right)=\delta_\iA^\iC\delta_\iB^\iD\;.
\eeq
From (\ref{wtwists}) and properties of matrices $F$ and $H$ we derive the following periodicity properties of the elliptic $r-$matrix (\ref{rmat}):
\beqn
\stackrel{12}{r}(\g+1)&=\stackrel1{F}\!^{-1}\;\stackrel{12}{r}\!\!(\g)\stackrel1{F}\;,
\non
\stackrel{12}{r}(\g+\mu)&=\stackrel1{H}\;\stackrel{12}{r}\!\!(\g)\stackrel1{H}\!^{-1}\;.
\label{bundle}
\eeqn
In the sequel we shall also need the functions
\beqn
\zi_{\iA\iB}(\g)=\frac{w_{\iA\iB}(\g)}{2\pi i}\Big(\frac{\theta_{[\iA\iB]}^\prime (\g)}{\theta_{[\iA\iB]}(\g)}-\frac{\theta_{[\iA\iB]}^\prime (0)}{\theta_{[\iA\iB]}(0)}\Big)\;,&(A,B)\neq (0,0)\;,
\label{zfunctions}
\eeqn
which have no singularities and transform as follows:
\beqn
\zi_{\iA\iB}(\g+1)=\epsilon^\iA\zi_{\iA\iB}\;,&\zi_{\iA\iB}(\g+\mu)=\epsilon^\iB\left(\zi_{\iA\iB}(\g)-w_{\iA\iB}(\g)\right)\;.
\label{zitwists}
\eeqn
Using the fact that theta functions satisfy the heat equation,
\beq
\frac{\partial^2\theta[p,q](\g;\mu)}{\partial \g^2}=4\pi i\frac{\partial\theta[p,q](\g,\mu)}{\partial\mu}\;,
\label{heat}
\eeq
we get the 
the following relation between $\zi_{\iA\iB}$ and $w_{\iA\iB}:$
\beq
\partial_\mu w_{\iA\iB}(\g;\mu)=\partial_\g\zi_{\iA\iB}(\g;\mu)\;.
\label{der_link}
\eeq
Now we are in a position to write down an ``elliptic'' counterpart of the linear
system (\ref{Psisystem}):
\beq
\frac{d\stackrel1{\Psi}(P)}{d\l_m}=\,\trt\Big(\rot(\nu(P)-\g_m)\stackrel2{J_m}\Big)\stackrel1{\Psi}(P)\;,
\label{rsystem}
\eeq
where 
\beqs
J_m=\,\sumablim J_m^{\iA\iB}\sigma_{\iA\iB}\;
\eeqs
with scalars $J_m^{\iA\iB}\;.$
Here $m=1,\dots,2N\;,$ $\Psi=\Psi(P,\{\l_m\})$ is a matrix-valued function; as before, $\nu(P)$ is the Abel map (\ref{abel}) from the covering $\L$ onto its fundamental domain $T=\C/\{1,\mu\}\;;$ $\g_m=\nu(P_m)\;.$ The compatibility condition of this system 
\beqn
\Big(\trt\Big(\rot(\nu(P)-\gm)\stackrel2{J_m}\Big)\Big)_{\l_n}-\Big(\trt\Big(\stackrel{12}{r}(\nu(P)-\gn)\stackrel2{J_n}\Big)\Big)_{\l_m}\non
+\Big[\trt\Big(\rot(\nu(P)-\gm)\stackrel2{J_m}\Big) , 
\trt\Big(\rot(\nu(P)-\gn)\stackrel2{J_n}\Big)\Big]=\,0
\label{compatibility}
\eeqn
gives the system of
differential equations for matrices $J_m$ as functions of the branch points $\l_m\;:$
\beqn
\frac{\d\Jmo}{\d\l_n}&=&\!-\;\;\alpha_n\Jmo\rho^\prime(\g_m-\g_n)\;-\;\alpha_m\trt\Big(\stackrel{12}{r}\!^\prime(\gm-\gn)\stackrel2{J_n}\Big)\;
\non
&-&\Big[
\Jmo ,\trt\Big(\rot(\gm-\gn)\stackrel2{J_n}\Big)\Big]\;,\hspace{0.4cm}m\neq n\;,
\label{jmsystem}
\eeqn
where $r^\prime$ stands for the derivative of the $r-$matrix with respect to its argument.
To prove that the compatibility condition reduces to (\ref{jmsystem}) we, first, compute the derivatives in (\ref{compatibility}) using the chain rule:
\beqs
r_{\l_n}(\g)=r_{\mu}(\g)\mu_{\l_n}+r^\prime(\g)\g_{\l_n}\;.
\eeqs
The derivative of the period $\mu$ is given by (\ref{mupolambdam}); for differentiation of $\nu$ and $\{\gm\}$ one uses the equations (\ref{glmint}) and (\ref{gnpolm}) respectively.
Then we note that the vector bundle $\chi$ over the Riemann surface $\L\;,$ whose monodromy matrices along the cycles $a$ and $b$ are given by $F^{-1}$ and $H$ respectively, is stable \cite{Hurtubise}. Checking the periodicity properties of the left hand side of (\ref{compatibility}) we see that it is a section of the adjoint bundle ${\mathrm{ad}}\chi\;.$ Due to the stability of $\chi$ the bundle  ${\mathrm{ad}}\chi$ 
does not have holomorphic sections (see for example \cite{Tyurin}). Therefore, for condition (\ref{compatibility}) to hold it suffices that the left hand side has no singularities; this is equivalent to the system (\ref{jmsystem}).

Equations (\ref{jmsystem}) form the non-autonomous non-linear integrable system associated with the space of elliptic coverings which gives an elliptic analog of the integrable system (\ref{hieint}).

\subsection{Tau-function }
\label{tau}
Let us introduce an object which we shall call the tau-function of the 
system (\ref{jmsystem}):
\beq
\frac{\d\log\tau}{\d\l_m}=\,\frac{1}{2\alpha_m}\mathrm{tr}(J_m^2)\;.
\label{taudef}
\eeq
To prove consistency of the definition we compute the derivatives of the right hand side, $\frac{\d}{\d\l_n}(\frac{1}{2\alpha_m}\mathrm{tr}(J_m^2))\;,$ using (\ref{jmsystem}). Then we get:
\beqs
\frac{\d^2\log\tau}{\d\l_m\d\l_n}=\,-\tro\;\trt\Big(\stackrel1{J_m}\stackrel2{J_n}\stackrel{12}{r}\!^\prime(
\gm-\gn)\Big)\;;
\eeqs
this expression is symmetric in $m$ and $n,$ due to the following properties of 
$r-$matrix:
$$\rot(\g)=\,-\rto(-\g)\;,$$ and  $$\stackrel{12}{r}\!^\prime(\g)=\,\stackrel{21}{r}\!^\prime(-\g).$$
This proves compatibility of the equations (\ref{taudef}).

An alternative definition of the tau-function (\ref{taudef}) can be given in terms of the 
one form $d\Psi\Psi^{-1}=\Psi_\nu\Psi^{-1}d\nu\;:$
\beq
\frac{\d\log\tau}{\d\l_m}=\frac{1}{2}{\mathrm{res}}|_{P_m}\Big\{\frac{\mathrm{tr}(d\Psi\Psi^{-1})^2}{d\l}\Big\}\;.
\label{taunewdef}
\eeq
To prove the equivalence of the two definitions, first note that we can write:
\beq
d\l=\frac{\d\l}{\d\nu}d\nu\;.
\eeq
Therefore using (\ref{glint}) for $\d\nu/\d\l$ we get
\beq
\frac{\mathrm{tr}(d\Psi\Psi^{-1})^2}{d\l}=\frac{\d\nu}{\d\l}\tr\left(\Psi_\nu\Psi^{-1}\right)^2 d\nu=\Big(\sum_{k=\,1}^{2N} \a_k\left(\rho(\nu-\g_k)+\rho(\g_k)\right)\Big)\tr(\Psi_\nu\Psi^{-1})^2d\nu\;.
\eeq
Further, we write the ``full" derivative of $\Psi$ with respect to $\l_m$ as follows:
\beq
\frac{d\Psi}{d\l_m}=\frac{\d\Psi}{\d\l_m}+\frac{\d\nu}{\d\l_m}\frac{\d\Psi}{\d\nu}\;,
\label{full}
\eeq
then using the form of the linear system (\ref{rsystem}) and formula (\ref{glmint}) for the derivative of $\nu\;,$ we rewrite (\ref{full}) in the form:
\beq
\trt\Big(\rot(\nu-\g_m)\stackrel2{J_m}\Big)=\frac{\d\stackrel1{\Psi}}{\d\l_m}\stackrel1{\Psi}\!^{-1}-\a_m\Big(\rho(\nu-\g_m)+\rho(\g_m)\Big)\frac{\d\stackrel1{\Psi}}{\d\nu}\stackrel1{\Psi}\!^{-1}\;,
\eeq
from which one can find $\tr(\Psi_\nu\Psi^{-1})^2$ and see that (\ref{taunewdef}) is equivalent to (\ref{taudef}). 

\subsection {Integrable system in the case of two-fold elliptic coverings}
The simplest elliptic covering $\L$ has two sheets and four ramification points. It corresponds to the hyperelliptic curve given by the following equation:
\beqs
\omega^2=(\l-\l_1)(\l-\l_2)(\l-\l_3)(\l-\l_4)\;;
\eeqs
$\l_m\;,\;\;m=1,\dots,4$ are branch points. On the covering we choose the basic cycle $a$ to encircle ramification points $P_1\;,\;P_2\;,$ and $b-$cycle to encircle points $P_2$ and $P_3\;.$ For this Riemann surface the normalized holomorphic differential ${\bf v}$ is given by 
\beq
{\bf v}=\frac{d\l}{\omega}\left[\oint_a\frac{d\l}{\omega}\right]^{-1}\;.
\label{hyperdiff}
\eeq
As before, $\mu$ is the $b-$period of the surface $\L\;:$ $\mu=\oint_b{\bf v}(P)\;.$ Consider the map $\tilde{\nu}$ from the covering $\L$ onto its fundamental domain $T=\C/\{1,\mu\}\;:$
\beqs
\tilde{\nu}(P)=\int_{P_1}^P{\bf v}(P)\;;
\eeqs
this map differs from the map $\nu(P)$ (\ref{abel}) by a function of branch points:
\beqs
\nu(P)=\tilde{\nu}(P)+h(\{\l_m\})\;,
\eeqs
where $h(\{\l_m\})=\int^{P_1}_{\infty^{(0)}}{\bf v}(P)\;.$ For our choice of basic cycles the images $\tilde{\g}_m$ of ramification points under the map $\tilde{\nu}$ are given by: 
\beqn
\tilde{\g}_1=\tilde{\nu}(P_1)=0\;;&\tilde{\g}_2=\tilde{\nu}(P_2)=\frac{1}{2}\;;
\non\nonumber
\tilde{\g}_3=\tilde{\nu}(P_3)=\frac{1}{2}+\frac{\mu}{2}\;;&\tilde{\g}_4=\tilde{\nu}(P_4)=\frac{\mu}{2}\;;
\eeqn
Since $\g_m-\g_n=\tilde{\g}_m-\tilde{\g}_n$ (where $\g_m=\nu(P_m)\;,\;m=1,\dots,4$ are as before the images of ramification points under the map $\nu$ (\ref{abel})), we can use these values of $\{\gt_m\}$ to write explicitly the system (\ref{jmsystem}) for the simplest covering. To do this we also calculate the coefficients
$\{\alpha_m\}_{m=1}^4$ defined by (\ref{amdef}). The form (\ref{hyperdiff}) of the normalized holomorphic differential ${\bf v}$ implies:
\beqs
\alpha_1=\frac{v_1^2}{2}=\frac{2}{(\l_1-\l_2)(\l_1-\l_3)(\l_1-\l_4)A^2}\;,
\eeqs
where $A=\oint_a\frac{d\l}{\omega}\;.$ From the Thomae formulae \cite{Fay92} we see that
\beqs
A^2=\frac{4\pi^2\theta_4^4}{(\l_1-\l_4)(\l_3-\l_2)}\;,
\eeqs
and therefore we have the following expressions for the coefficients $\alpha_m\;:$
\beqn
\alpha_1=\frac{\l_3-\l_2}{2\pi^2\theta_4^4(\l_1-\l_2)(\l_1-\l_3)}\;;&\alpha_2=-&\frac{\l_1-\l_4}{2\pi^2\theta_4^4(\l_2-\l_1)(\l_2-\l_4)}\;;
\non\nonumber
\alpha_3=\frac{\l_1-\l_4}{2\pi^2\theta_4^4(\l_3-\l_1)(\l_3-\l_4)}\;;&\alpha_4=-&\frac{\l_3-\l_2}{2\pi^2\theta_4^4(\l_4-\l_2)(\l_4-\l_3)}\;.
\eeqn
Now we can write down the integrable system (\ref{jmsystem}) explicitly for $K=2$ ($K$ is the matrix dimension of the system). In this case we use the standard Pauli basis $\{\sigma_1\;,\;\sigma_2\;,\;\sigma_3\}$ related to the matrices $\sigma_{\iA\iB}$ as follows:
\beqn
\sigma_{10}=\sigma_1\;,&\sigma_{11}=i\,\sigma_2\;,&\sigma_{01}=\sigma_3\;;
\non
\sigma^{10}=\frac{1}{2}\sigma_1\;,&\sigma^{11}=\frac{i}{2}\sigma_2\;,&\sigma^{01}=\frac{1}{2}\sigma_3\;.
\label{pauli}
\eeqn
The corresponding notation for components of $J_m$ is:
\beqn
J_m^1=J_m^{10}\;,&J_m^2=iJ_m^{11}\;,&J_m^3=J_m^{01}\;.
\label{jms}
\eeqn
We shall
write the equations for $(J_1)_{\l_2}$  ($J_1=J_1^1\sigma_1+J_1^2\sigma_2+J_1^3\sigma_3$). The remaining equations for $(J_m)_{\l_n}$  in the case of two-fold elliptic covering have a similar form. 
\beqn
\frac{\d J_1^1}{\d\l_2}&=&\frac{\l_1-\l_4}{2\pi^2(\l_2-\l_1)(\l_2-\l_4)}J_1^1\frac{1}{\theta_4^4}\frac{\theta^{\prime\prime}_2}{\theta_2}+2\pi i J_1^3J_2^2\theta_4^2\;,\nonumber\\
\frac{\d J_1^2}{\d\l_2}&=&\frac{\l_1-\l_4}{2\pi^2(\l_2-\l_1)(\l_2-\l_4)}J_1^2\frac{1}{\theta_4^4}\frac{\theta^{\prime\prime}_2}{\theta_2}-2\pi i J_1^3J_2^1\theta_3^2\;,\nonumber\\
\frac{\d J_1^3}{\d\l_2}&=&\frac{\l_1-\l_4}{2\pi^2(\l_2-\l_1)(\l_2-\l_4)}J_1^3\frac{1}{\theta_4^4}\frac{\theta^{\prime\prime}_2}{\theta_2}+\frac{\l_3-\l_2}{2(\l_1-\l_2)(\l_1-\l_3)}J_2^3\frac{\theta_3^2}{\theta_4^2}\nonumber\\
&+&2\pi i\left(J_1^2J_2^1\theta_3^2-J_1^1J_2^2\theta_4^2\right)\;.
\nonumber
\eeqn
Here $\theta_2=\theta[\frac{1}{2},0](0)\;;\;\theta_3=\theta[0,0](0)\;;\;\theta_4=\theta[0,\frac{1}{2}](0)$ and $\theta_2^{\prime\prime}=\theta_2^{\prime\prime}(0)$ are the standard theta-constants.

\section{Relationship to the Schlesinger system}
\label{Schles}
The elliptic Schlesinger system \cite{Takasaki} describes isomonodromic deformations of 
solutions $\Psi(\g,\{z_i\})$ of the following matrix linear differential equation:
\beq
\frac{d\Psi}{d\g}=\,A(\g)\Psi,
\label{psiequation}
\eeq
where $\g$ is a coordinate on the torus $T=\C/\{1,\mu\}\;$; $A(\gamma)$ is a meromorphic $sl(K,\C)-$valued matrix:
\beqs
\stackrel1{\!\!A}(\g)=\,\sum_{j=\,1}^L\trt\Big(\rot(\g-z_j)\stackrel{2}{A_j}\Big);\,
\eeqs
 $r(\g)$ is the elliptic $r-$matrix (\ref{rmat});
$z_j\in T\;,\;j=1,\dots,L\;;$ $L$ is some integer. At the points $\{z_j\}$ the matrix $A(\g)$ has simple poles with residues $A_j\;$. The residues are, in turn, parameterized as follows:
\beq
A_j=\sumablim A_j^{\iA\iB}\sigma_{\iA\iB}\;,
\label{Acomponents}
\eeq
where matrices $\sigma_{\iA\iB}$ are given by (\ref{sigmab}); $A_j^{\iA\iB}\in\C\;.$
The matrix $A(\gamma)$ has the following periodicity properties:
\beqs
A(\gamma+1)=F^{-1}A(\gamma)F\;,\hsp A(\gamma+\mu)=HA(\gamma)H^{-1}\;.
\eeqs
It is assumed that $\Psi$ has asymptotical expansion near $z_j\;,$ $j=1,\dots,L\;,$ of the form:
\beq
\Psi(\g)=\left(G_j+O(\g-z_j)\right)(\g-z_j)^{T_j}C_j\;,
\label{psiatlambdaj}
\eeq
where matrices $G_j\;,C_j\;,T_j$ do not depend on $\g\,;\;C_j\;,G_j\in SL(K,\C)$ and 
$T_j$ are diagonal traceless matrices such that any two entries of $T_j$ do not differ by an integer number. 
The function $\Psi$ transforms as follows with respect to periods $1$ and $\mu$ of the torus $T$: 
\beqn
\Psi(\g+1)&=&F^{-1}\Psi(\g)M_a\;,
\non
\Psi(\g+\mu)&=&H\Psi(\g)M_b\;,
\nonumber
\eeqn
and being analytically continued along a contour $l_j$ surrounding the point $z_j\;$ the function $\Psi$ gains a right multiplier:
\beqn
\nonumber
\Psi(\g^{l_j})&=&\Psi(\g)M_j\;,
\eeqn
where $M_a\;,M_b\;,M_j$ are called the monodromy matrices. 
The assumption of independence of all monodromy matrices of the positions of singularities $\{z_j\}$ and the $b-$period $\mu$ of the elliptic Riemann surface is called the isomonodromy condition. This condition together with expansion (\ref{psiatlambdaj}) gives 
the following dependence of $\Psi$ on $\mu$ and $\{z_j\}_{j=1}^L$:
\beq
\stackrel1{\Psi}_{z_i}\stackrel1{\Psi}\!^{-1}=\,-\trt\Big(\rot(\g-z_i)\stackrel2{A_i}\Big)\;,
\label{psipomu}
\eeq
\beq
\Psi_\mu\Psi^{-1}=\,\sum_{j=\,1}^L\sumablim 
A_j^{\iA\iB}\zi_{\iA\iB}(\g-z_i)\sigma_{\iA\iB}\;,
\label{psipozj}
\eeq
the functions $\zi_{\iA\iB}$ were defined by (\ref{zfunctions}).
The compatibility condition of (\ref{psipomu}), (\ref{psipozj}) and 
(\ref{psiequation}) gives the Schlesinger system on the elliptic surface:
\beqn
\frac{\d\stackrel1{A_i}}{\d 
z_j}&=&\,\Big[\stackrel1{A_i},\trt\Big(\rot(z_i-z_j)\stackrel2{A_j}\Big)\Big]\;,\hspace{0.2cm}i\neq j\;,
\non
\frac{\d\stackrel1{A_i}}{\d z_i}&=&\!\!-\sum_{j=1,j\neq i}^L\Big[\stackrel1{A_i},\trt\Big(\rot(z_i-z_j)\stackrel2{A_j}\Big)\Big]\;,
\non
\frac{\d\stackrel1{A_i}}{\d 
\mu}&=&\!\!-\sum_{j=1}^L\Big[\stackrel1{A_i},\trt\Big(\stackrel2{A_j}\!\!\!\sumablim\!\!\!\zi_{\iA\iB}(z_i-z_j)\stackrel1{\sigma}_{\iA\iB}
\stackrel2{\sigma}\!^{\iA\iB}\Big)\Big]\;.
\label{Schlesinger}
\eeqn
The tau-function of this system is defined as the generating function of the following Hamiltonians:
\begin{eqnarray}
H_i\!\!\!&\!\!=\!\!\!&\,\frac{1}{4\pi i}\oint_{z_i}trA^2(\g)d\g=\,\sum_{j=1,j\neq i}^L\sumablim\!\!\! 
A_j^{\iA\iB}A_{i\iA\iB}w_{\iA\iB}(z_i-z_j)\;,\label{hamiltonianj}\\
H_\mu\!\!\!&\!\!=&\!\!\!\,-\frac{1}{2\pi i}\oint_a 
trA^2(\g)d\g=\,\frac{1}{2}\sum_{i,j=1}^L\sumablim\!\!\!\!
A_j^{\iA\iB}A_{i\iA\iB}\zi_{\iA\iB}(z_i-z_j)\;.\label{hamiltonianmu}
\end{eqnarray}
\beq
\frac{\d\log\tau_{\mathrm Sch}}{\d z_i}=H_i\;,\hsp\frac{\d\log\tau_{\mathrm Sch}}{\d\mu}=H_\mu\;.
\label{tauSchlesinger}
\eeq
The following theorem shows how (analogously to the rational case \cite{KokKor}) solutions of the elliptic Schlesinger system (\ref{Schlesinger}) induce solutions of system (\ref{rsystem}) and (\ref{jmsystem}).
\begin{theorem}
Let $\L$ be a genus $1$ covering of the $\l-$sphere with simple 
ramification points $P_1,\dots,P_{2N}\;,$ which have different $\l-$projections $\l_1,\dots,\l_{2N}\;.$ Consider a set of $L$ points $\{Q_1,\dots,Q_L\}$ 
on $\L$ such that their projections $\pi(Q_i)$ are independent of $\{\l_m\}\;$. Let $\nu$ be the Abel map (\ref{abel}) onto the fundamental domain of the covering, $\nu:\L\rightarrow T\;.$ Consider the Schlesinger system 
(\ref{Schlesinger}) with $z_i=\,\nu(Q_i)$ and its solution 
$\{A_j(\{z_i\})\}_{j=1}^L\;.$  Let $\Psi(\g,\{z_i\})$ be the corresponding solution 
of system (\ref{psiequation}). We can consider $\Psi$ as a function on the covering
$\L$ via the Abel map:
\beq
\Psi(P)=\,\Psi(\nu(P),\{\nu(Q_i)\}).
\label{psi}
\eeq
Then
\begin{enumerate}
\item the function $\Psi(P)$ 
satisfies the linear system (\ref{rsystem}) with $J_m$ defined by
\beq
\stackrel1{J_m}=\,-\alpha_m\sum_{j=1}^L\trt\Big(\rot(\gm-z_j)\stackrel2{A_j}\Big)\;,
\label{Jms}
\eeq
and, hence, $J_m$'s solve the system (\ref{jmsystem});
\item the tau-function $\tau$ (\ref{taudef}) of the system (\ref{jmsystem}) is related to the tau-function $\tau_{\mathrm Sch}$ (\ref{tauSchlesinger}) of the elliptic Schlesinger system according to:
\beq
\tau(\l_m)=\,\prod_{j=\,1}^L\Big(\frac{\d\nu}{\d\l}(Q_j)\Big)^{\tr
A^2_j/2}\tau_{\mathrm Sch}\Big(\{z_k\}_{|_{z_k=\,\nu(Q_k)}}\Big)\;.
\label{tautau}
\eeq
\end{enumerate}
\end{theorem}
\textbf{Remark}. Formula (\ref{tautau}) coincides with the one relating the tau-function of the rational system (\ref{Psisystem}) and the tau-function of the Schlesinger system on the Riemann sphere, see \cite{KokKor}.

\vspace{0,5cm}

{\it{Proof.}} Since the solution $\Psi$ of (\ref{psiequation}) is defined on the space of branch coverings as in (\ref{psi}), we can
differentiate it with respect to $\l_m$ according to the chain rule:
\beqs
\frac{d\Psi}{d\l_m}=\,\frac{\d\Psi}{\d\nu}\frac{\d\nu}{\d\l_m}+\sum_{j=1}^L\frac{\d\Psi}{\d z_j}\frac{\d z_j}{\d\l_m}+\frac{\d\Psi}{\d\mu}\frac{\d\mu}{\d\l_m}\;.
\eeqs
(Recall that $\mu$ is the $b-$period of the elliptic Riemann surface.)
We differentiate $z_i=\nu(Q_i)$ according to the formula (\ref{glmint}) for derivatives of $\nu$ and use also formulae (\ref{mupolambdam}),  (\ref{psiequation}), (\ref{psipomu}), (\ref{psipozj}). 
Using the relation
\beqn
\nonumber
{w_{\iA\iB}(\g-z_j)}\Big(\rho(z_j-\gm)-\rho(\g-\gm)\Big)+2\pi i\zi_{\iA\iB}(\g-z_j)\\
\nonumber
=\,-{w_{\iA\iB}(\gm-z_j)}w_{\iA\iB}(\g-\gm)\;,
\eeqn
which can be proved by checking periodicity properties of both sides as 
$\g\rightarrow\g+1,$ $\g\rightarrow\g+\mu$ and behavior at the pole 
$\g=\gm\;,$ we obtain:
\beq
\frac{d\stackrel1{\Psi}}{d\l_m}=-\alpha_m\stackrel2{\tr}\Big(\stackrel{12}{r}(\nu-\g_m)\sum_{j=1}^{L}\stackrel3{\tr}\Big(\stackrel{23}{r}(\g_m-z_j)\stackrel3{A}_j\Big)\Big)\stackrel1{\Psi}\;.
\label{tempo}
\eeq
We single out the $\nu-$dependent term and denote the rest by $J_m$:
\beqn
\nonumber
\stackrel2{J_m}:
=\,-\alpha_m\sum_{j=1}^{L}\stackrel3{\tr}\Big(\stackrel{23}{r}(\gm-z_j)\stackrel3{A_j}\Big)\;.
\eeqn
This leads to the system
(\ref{rsystem}) and proves the first part of the theorem. For 
the second part, equality (\ref{tautau}), we shall prove the following relation between the two tau-functions:
\beq
\frac{\d\log\tau}{\d\l_m}=\,\frac{\d\log\tau_{\mathrm Sch}}{\d\l_m}+\sum_{j=1}^{L}\frac{\tr A_j^2}{2}\frac{\d\log\frac{\d\nu}{\d\l}}{\d\l_m}\vert_{\nu=\,z_j}\;.
\label{tautautau}
\eeq
This leads to (\ref{tautau}) if one observes that
\beqs
\frac{\d\;\tr A_j^2}{\d\l_m}=0\;,
\eeqs
which follows from the Schlesinger system (\ref{Schlesinger}).
To show (\ref{tautautau}) let us first note two auxiliary relations. The first one is:
\beqn
\nonumber
&&w_{\iA\iB}(z_i-z_j)\Big(\rho(z_j-\g)-\rho(z_i-\g)\Big)\\
&=&\,w_{\iA\iB}(\g-z_j) w_{-\iA-\iB}(\g-z_i)-2\pi i\;\zi_{\iA\iB}(z_i-z_j)\;
\label{auxil-1}
\eeqn
for any pair of non equal indeces $i\;,\;j\;.$
This relation can be verified examining singularities and periodicity properties and then noting that at the point $\g=\,\frac{1}{2}(z_i+z_j)$ both sides 
are equal due to the equality
\beqs
2w_{\iA\iB}(2\g)\rho(\g)=\,w_{\iA\iB}^2(\g)+2\pi 
i\zi_{\iA\iB}(2\g),
\eeqs
which, in turn, can be proved by the same method. One can apply the similar considerations to verify the second identity which we shall use:
\beq
w_{\iA\iB}(\g)w_{-\iA-\iB}(\g)=2\pi i\zi_{\iA\iB}(0)-\rho^\prime(\g)\;.
\label{auxil-2}
\eeq
To show (\ref{tautautau}) we differentiate the tau-function of the elliptic Schlesinger system $\tau_{\mathrm Sch}$ with respect to $\l_m:$
\beqn
\nonumber
\frac{\d\log\tau_{\mathrm Sch}}{\d\l_m}=\,\sum_{i=1}^L\frac{\d\log\tau_{\mathrm Sch}}{\d z_i}\frac{\d z_i}{\d\l_m}+\frac{\d\log\tau_{\mathrm Sch}}{\d\mu}\frac{\d\mu}{\d\l_m}\;.
\label{tempo1}
\eeqn
Then we rewrite all the terms explicitly using (\ref{tauSchlesinger}), (\ref{hamiltonianj}), (\ref{hamiltonianmu}), (\ref{glmint}), (\ref{mupolambdam}) and simplify the obtained expression applying the auxiliary identity (\ref{auxil-1}). Noting also that
\beqn
\zi_{-\iA-\iB}(-\g)=\,\zi_{\iA\iB}(\g)\;, 
\eeqn
one arrives at the following expression:
\beqn
\nonumber
\frac{\d\log\tau_{\mathrm Sch}}{\d\l_m}\!\!\!&\!\!\!=&\!\!\!\alpha_m K\!\Big(\sum_{\stackrel{i,j=1}{i<j}}^L\!\sumablim\!\!\!\!\!\epsilon^{\iA\iB}A_j^{\iA\iB}A_i^{-\iA-\iB} 
w_{\iA\iB}(\gm-z_j)w_{-\iA-\iB}(\gm-z_i)\Big.\\
&+&\Big.\pi i\sum_{i=1}^L\sumablim\epsilon^{\iA\iB}A_i^{\iA\iB}A_i^{-\iA-\iB}\zi_{\iA\iB}(0)\Big).
\label{1term}
\eeqn
The derivative in the second term of the right hand side of (\ref{tautautau}) can be obtained using (\ref{glmint}) as follows:
\beqn
\nonumber
\frac{\d\log\frac{\d\nu}{\d\l}}{\d\l_m}=\frac{\d}{\d\l}\Big(\frac{\d\nu}{\d\l_m}\Big){\Big/}\frac{\d\nu}{\d\l}
=\frac{\d}{\d\nu}\Big(\frac{\d\nu}{\d\l_m}\Big)\;,
\eeqn
hence
\beq
\frac{\d\log\frac{\d\nu}{\d\l}}{\d\l_m}\vert_{\nu=z_j}=-\alpha_m\rho^\prime(z_j-\gm)\;.
\eeq
A certain simplification using the second auxiliary identity (\ref{auxil-2}) leads to the following expression for the right hand side of (\ref{tautautau}):
\beqn
\nonumber
&&\frac{\d\log\tau_{\mathrm Sch}}{\d\l_m}+\sum_{i=1}^L\frac{\tr A_i^2}{2}\frac{\d\log\frac{\d\nu}{\d\l}}{\d\l_m}\vert_{\nu=\,z_i}\\
&\!\!\!=\!\!\!&\!\!\!\frac{\alpha_m}{2}\!\!\Big(\!\sum_{i,j=1}^L\!\!\sumablim\!\!\!\!\!\!\!\!K\epsilon^{\iA\iB}\!A_j^{\iA\iB}A_i^{-\iA-\iB} 
w_{\iA\iB}(\gm-z_j)w_{-\iA-\iB}(\gm-z_i)\!\Big)\;,
\eeqn
which is nothing but $\tr(J^2_m)/{2\alpha_m}\;$, where $J_m$ are given by (\ref{Jms}). Thus the right hand side of (\ref{tautautau}) is equal to $(\log\tau)_{\l_m}$ where the tau-function $\tau$ is defined by (\ref{taudef}).

$\Box$

\section{Trigonometric degeneration of the elliptic coverings and corresponding integrable systems.}
\label{example}
Here we describe the trigonometric version of system (\ref{jmsystem}), obtained by a degeneration of the covering $\L.$ Further, 
as an illustration, we consider the two-fold covering when all coefficients of the obtained system   can be computed explicitely. 

Set the matrix dimension $K$ of the system to be $2\;$. An elliptic $N-$fold covering has, according to the Riemann-Hurwitz formula, $2N$ branch points (recall that we assume them to be simple and distinct). If we let one branch cut to degenerate (i.e.,  we let two ramification points connected by a branch cut to tend to each other), the elliptic covering turns into a rational one with $2N-2$ ramification points and a double point remaining from the degenerated branch cut. 

Assume that the points
$P_{2N-1}$ and $P_{2N}$ are connected by a  branch cut $[P_{2N-1}, P_{2N}]$.
Moreover, choose the  
basic $a$-cycle on $\L$ to surround this branch cut.
Consider  $\{\l_m\}_{m=1}^{2N-2}$ as independent variables and 
$\l_{2N-1}$ and $\l_{2N}$ as fixed parameters. Take the limit
$\l_{2N-1}, \l_{2N}\to \l_Q$ with $\l_Q$ independent of $\{\l_m\}_{m=1}^{2N-2}$. Then the
branch cut $[P_{2N-1}, P_{2N}]$ degenerates and the
elliptic curve $\L$  turns into the rational curve $\L_0$
with two marked points $Q_1$ and $Q_2$ (a double point) which lie on
different sheets of $\L_0$ and have the same projection on the
$\l$-plane:
$$\pi(Q_1)=\pi(Q_2)=\l_Q\;.$$ 
The basic $a$-cycle on $\L$ turns into a contour on $\L_0$ surrounding 
one of the points $Q_1$ or $Q_2$. Suppose that it surrounds $Q_1$ in the
positive direction.
Denote by $\zeta(P),\;P\in\L_0$ the one-to-one map from the genus zero
covering $\L_0$  with ramification points $P_1,\dots,P_{2N-2}$ to the
Riemann sphere; for simplicity we fix this map by requirement
$\zeta:\infty^{(0)}\to \infty$ such that in a neighborhood of $\infty^{(0)}$ 
\beq
\zeta(\l)=\l+o(1)\;.
\la{normuni}
\eeq
Denote the images of points $Q_1$ and $Q_2$ on the Riemann sphere by $\kappa_1$ and $\kappa_2$ respectively:
\beq
\kappa_1=\zeta(Q_1)\;;\hskip0.7cm 
\kappa_2=\zeta(Q_2)\;.
\label{kappy}
\eeq
The holomorphic differential ${\bf v}(P)$ degenerates to the meromorphic on $\L_0$
differential ${\bf v}_0$ with the simple poles at $Q_1$ and $Q_2$ and
residues $1/2\pi i$ and $-1/2\pi i$ respectively. 
This differential can be written in terms of the coordinate $\zeta$
as follows:
\beq
{\bf v}_0(\zeta) = \f{1}{2\pi i}\Big(\f{1}{\zeta-\kappa_1}-\f{1}{\zeta-\kappa_2}\Big)d\zeta\;.
\label{differential}
\eeq
The $b$-period $\mu$ of the Riemann surface $\L$ in the limit $P_{2N-1}\to P_{2N}$ has the following behavior:
\beq
\mu= \f{1}{\pi i}\log |\l_{2N-1} - \l_{2N}| +O(1)\;,
\eeq
i.e., $\mu\to + i\infty$ in this limit, and the fundamental domain $T=\C/\{1,\mu\}$ of the covering $\L$ turns into a cylinder. The map $\nu$ (\ref{abel}) now maps the degenerated covering $\L_0$ onto the cylinder in $\g-$sphere:
\beq
\nu(P)=\int_{\infty^{(0)}}^P {\bf v}_0 = \f{1}{2\pi
i}\int_{\infty}^{\zeta}\left(\f{1}{\zeta-\kappa_1}-\f{1}{\zeta-\kappa_2}\right)d\zeta
=\f{1}{2\pi i}\log\f{\zeta-\kappa_1}{\zeta-\kappa_2}\;.
\label{gtri}
\eeq
From the definition (\ref{thetadef}) of the
Jacobi theta-functions,
we deduce the behavior of logarithmic derivative $\rho(\g)$ of $\theta_1=\theta[\frac{1}{2},\frac{1}{2}]$ as $\mu\to+i\infty$ :
\beq
\rho(\g)\to\pi\cot\pi\g\;,
\eeq
and therefore, 
\beq
\rho'(\g) \to -\f{\pi^2}{\sin^2\pi\g}\;.
\eeq
Similarly, the $r-$matrix becomes in this limit (for the matrix dimension $K=2$):
\beq
\stackrel{12}{r}(\gamma)\to\stackrel{12}{r}\!\!\!_0(\g)=\frac{1}{2}\frac{\pi}{\sin\pi\gamma}\stackrel1{\sigma}\!_1\!\stackrel2{\sigma}\!_1+\frac{1}{2}\frac{\pi}{\sin\pi\gamma}\stackrel1{\sigma}\!_2\stackrel2{\sigma}\!_2+\frac{1}{2}\pi\cot\pi\gamma\stackrel{1}{\sigma}_3\stackrel{2}{\sigma}\!_3\;,
\label{rlim}
\eeq
where we use the Pauli basis $\{\sigma_i\}_{i=1}^3$ (\ref{pauli}); $\;r_0$ is the so-called trigonometric $r-$matrix.
Differential equations (\ref{gnpolm})--(\ref{difmpolambdam}) for $\{\g_m\}_{m=1}^{2N-2}$ (images of non-de\-ge\-ne\-rated ramification points $P_1,\dots,P_{2N-2}$ under the map $\nu$ (\ref{gtri})) take the form (for $m\neq n$):
\beqn
\frac{\d\g_n}{\d\l_m}&=\!\!&\!\!\!-\pi\alpha_m^0\left[\cot\pi(\g_n-\g_m)+\cot\pi\g_m\right]\;,
\label{trigdiff}\\
\frac{\d\g_m}{\d\l_m}&=&\pi\!\!\sum_{n=1,\, n\neq 
m}^{2N-2}\alpha_n^0\left[\cot\pi(\g_m-\g_n)+\cot\pi\g_n\right]\;,
\non
\frac{\d\a_n^0}{\d\l_m}&=&2\,\pi^2\frac{\a_n^0\a_m^0}{\sin^2\pi(\g_n-\g_m)}\;,
\label{atrigdiff}\\
\frac{\d\a_m^0}{\d\l_m}&=\!\!&\!\!\!-2\,\pi^2\!\!\sum_{n=1,\, n\neq 
m}^{2N-2}\frac{\a_n^0\a_m^0}{\sin^2\pi(\g_n-\g_m)}\;,
\nonumber
\eeqn
where by $\a_m^0$ we denoted the analog of the coefficient $\a_m$ in the degenerated case:
\beq 
\a_m^0=\f{1}{2}v^2_{0m}= \f{1}{2}\Big[\frac{{\bf v_0}(P)}{d\sqrt{\l-\l_m}}\Big|_{P=P_m}\Big]^2\;,\hspace{0.4cm} m=1,\dots,2N-2\;.
\la{am0def}
\eeq

{\bf Remark 1.} Differential equations (\ref{trigdiff}) can be obtained directly from the form (\ref{gtri}) of the map $\nu$ using the fact that the map $\zeta$ satisfies equations (\ref{gammamdiff}) with $\beta=1\;,\; c_n=0\;.$

{\bf Remark 2.} The system (\ref{trigdiff}), (\ref{atrigdiff}) after a simple change of variables coincides with equations for characteristic speeds of the system of hydrodynamic type to which the Boyer-Finley equation (self-dual Einstein equation with one Killing vector) $U_{xy}=(e^U)_{tt}$ reduces \cite{FKS}.
\vspace{0,5cm}

The linear system for the matrix $\Psi$ is written now via the trigonometric $r-$matrix $r_0$: 
\beq
\frac{d\stackrel1{\Psi}(P)}{d\l_m}=\,\trt\Big(\rot_0(\nu(P)-\g_m)\stackrel2{J_m}\Big)\stackrel1{\Psi}(P)\;,
\label{r0system}
\eeq
$m=1,\dots,2N-2\;.$
Then, the trigonometric version of system (\ref{jmsystem}) for $J_m=J_m^1\sigma_1+J_m^2\sigma_2+J_m^3\sigma_3\;$ (for notation see (\ref{jms})) gives the compatibility condition of the above linear system: 
\beqn
\frac{\d J^1_m}{\d\l_n}&=&\frac{\alpha_n^0\pi^2}{\sin^2\pi(\g_m-\g_n)}J^1_m
+\frac{\alpha_m^0\pi^2\cos\pi(\gm-\gn)}{\sin^2\pi(\gm-\gn)}J_n^1
\non
&+&\frac{2\pi i}{\sin\pi(\g_m-\g_n)}\Big(J_m^2J_n^3\cos\pi(\g_m-\g_n)-J_m^3J_n^2\Big)\;,
\non
\frac{\d J^2_m}{\d\l_n}&=&\frac{\alpha_n^0\pi^2}{\sin^2\pi(\g_m-\g_n)}J^2_m
+\frac{\alpha_m^0\pi^2\cos\pi(\gm-\gn)}{\sin^2\pi(\gm-\gn)}J_n^2
\label{trigsys}\\
&+&\frac{2\pi i}{\sin\pi(\g_m-\g_n)}\Big(J_m^3J_n^1-J_m^1J_n^3\cos\pi(\g_m-\g_n)\Big)\;,
\non
\frac{\d J^3_m}{\d\l_n}&=&\frac{\alpha_n^0\pi^2}{\sin^2\pi(\g_m-\g_n)}J^3_m
+\frac{\alpha_m^0\pi^2}{\sin^2\pi(\gm-\gn)}J_n^3
\non
&+&\frac{2\pi i}{\sin\pi(\g_m-\g_n)}\Big(J_m^1J_n^2-J_m^2J_n^1\Big)\;,
\nonumber
\eeqn
the indeces $m$ and $n$ are different and range in the set $\{1,\dots,2N-2\}$.

All the involved coefficients can be explicitly computed if we start with the two-fold elliptic covering. After the degeneration we get a rational covering $\L_0$ with two ramification points $P_1$ and $P_2$ (with the $\l-$projections $\l_1$ and $\l_2$) and the marked points with the projection $\l_Q\;$ independent of $\l_1$ and $\l_2\;.$ The one-to-one map $\zeta$ from this covering to the Riemann sphere which satisfies condition (\ref{normuni}) has the following form:
\beq
\zeta(P)=\frac{1}{2}\Big(\lambda+\frac{\lambda_1+\lambda_2}{2}+\sqrt{(\lambda-\lambda_1)(\lambda-\lambda_2)}\Big)\;,
\label{twopointmap}
\eeq
where $\l=\pi(P)\;,$ the projection of the point $P$ on the base of the covering.
Knowing the expression for the map $\zeta$ allows us to find the images $\g_1\;,\g_2$ of the non-degenerate ramification points $P_1\;,P_2$ under the map $\nu(P)$  since it can be explicitly integrated (see(\ref{gtri})):
\beq
\g_m=\frac{1}{2\pi i}\log\f{\zeta_m-\kappa_1}{\zeta_m-\kappa_2}\;, 
\label{gmtri}
\eeq
where $\zeta_m=\zeta(P_m)\;,\; m=1,2$ are the images on the Riemann sphere of 
ramification points.
One can find them from the form (\ref{twopointmap}) of the map $\zeta\;:$ 
\beqs
\zeta_1=\zeta(\lambda_1)=\frac{3\lambda_1+\lambda_2}{4}\;,\hsp\zeta_2=\zeta(\lambda_2)=\frac{\lambda_1+3\lambda_2}{4}\;;\\
\eeqs
for the $\zeta-$images $\kappa_{1,2}$ of points $Q_1$ and $Q_2$ (\ref{kappy}) we have:
\beqs
\kappa_{1,2}=\frac{1}{2}\Big(\lambda_Q+\frac{\lambda_1+\lambda_2}{2}\pm\sqrt{(\lambda_Q-\lambda_1)(\lambda_Q-\lambda_2)}\Big)\;.
\eeqs
Now one can easily see from (\ref{gmtri}) that
\beq
e^{2\pi i\gamma_1}=\frac{\sqrt{\lambda_1-\lambda_Q}-\sqrt{\lambda_2-\lambda_Q}}{\sqrt{\lambda_1-\lambda_Q}+\sqrt{\lambda_2-\lambda_Q}}=-e^{2\pi i\gamma_2}\;,
\label{gammytrig}
\eeq
and, therefore, $\gamma_1-\gamma_2=\pm1/2$ . The same conclusion can be made if one observes that $\gamma_1-\gamma_2$ is equal to one half of the integral over $a-$period of the differential ${\bf v}$ (see definition (\ref{abel}) of the map $\nu$). The sign of the difference $\gamma_1-\gamma_2$ is determined by a choice of direction of the $a-$cycle. 

It remains to calculate one more ingredient of system (\ref{jmsystem}) for $J_m$, namely, the coefficients $\alpha^0_{1,2}$ (\ref{am0def}).
Denoting by $v_0(x)$ 
a locally defined function such that ${\bf v_0}=v_0(x)dx$ ($x$ being a local parameter on the covering), from the relation
\beqs
v_0(x)dx=\f{1}{2\pi i}\Big(\f{1}{\zeta-\kappa_1}-\f{1}{\zeta-\kappa_2}\Big)d\zeta\;
\eeqs
we deduce that 
\beqs
\frac{d\zeta}{dx}(\l_m)=2\pi i\;v_{0m}\frac{(\zeta_m-\kappa_1)(\zeta_m-\kappa_2)}{\kappa_1-\kappa_2}\;,\hspace{0.4cm}m=1,2\;.
\eeqs
From the explicit form (\ref{twopointmap}) of the map $\zeta(P)$ one can compute the coefficients $(d\zeta/dx)(\l_m)$ of expansion of $\zeta(P)$ in neighbourhoods of ramification points $P_1\;,\;P_2\;.$ Then we obtain the expressions for $\alpha^0_m=\frac{1}{2}v_{0m}^2\;\;(m=1,2)$ :
\beqs
\alpha^0_1=-\frac{1}{2\pi^2}\frac{\lambda_2-\lambda_Q}{\lambda_1-\lambda_Q}\frac{1}{\lambda_1-\lambda_2}\;,\hsp
\alpha^0_2=-\frac{1}{2\pi^2}\frac{\lambda_1-\lambda_Q}{\lambda_2-\lambda_Q}\frac{1}{\lambda_2-\lambda_1}\;.
\eeqs
In the limit $\l_Q\to\infty$ , summarizing all the above calculations, we get from (\ref{trigsys}) the system of equations for $J_1=J_1^1\sigma_1+J_1^2\sigma_2+J_1^3\sigma_3$ :
\beqn
\frac{\d J^1_1}{\d\l_2}&=&\frac{1}{2}\frac{1}{\l_1-\l_2}J^1_1-2\pi i\;J^3_1J^2_2\;,
\non
\frac{\d J^2_1}{\d\l_2}&=&\frac{1}{2}\frac{1}{\l_1-\l_2}J^2_1+2\pi i\;J^3_1J^1_2\;,
\non
\frac{\d J^3_1}{\d\l_2}&=&\frac{1}{2}\frac{1}{\l_1-\l_2}(J^3_1-J_2^3)+2\pi i\;(J^1_1J^2_2-J_1^2J_2^1)\;,
\nonumber
\eeqn
and the similar system for $J_2$ :
\beqn
\frac{\d J^1_2}{\d\l_1}&=&\frac{1}{2}\frac{1}{\l_2-\l_1}J^1_2+2\pi i\;J^3_2J^2_1\;,
\non
\frac{\d J^2_2}{\d\l_1}&=&\frac{1}{2}\frac{1}{\l_2-\l_1}J^2_2-2\pi i\;J^3_2J^1_1\;,
\non
\frac{\d J^3_2}{\d\l_1}&=&\frac{1}{2}\frac{1}{\l_1-\l_2}(J^3_1-J_2^3)+2\pi i\;(J^1_1J^2_2-J_1^2J_2^1)\;.
\nonumber
\eeqn
\vspace{0.3cm}

{\bf Acknowledgment.} I thank D.Korotkin for advising me in the course of the work and careful reading of the manuscript. I also thank E. Ferapontov, A. Kokotov and A. Mikhailov for useful discussions.


\begin{thebibliography}{99}
\bibitem{Takasaki}Takasaki, K., Gaudin model, KZ equation and an isomonodromic problem on the torus, Lett. Math. Phys. {\bf 44}, 143--156 (1998)
\bibitem{ZMNP} Novikov, S., Manakov, S. V., Pitaevski\u\i, L. P., Zakharov, V. E., Theory of solitons. The inverse scattering method, Contemporary Soviet Mathematics. Consultants Bureau [Plenum] (1984) (Translated from the Russian)
\bibitem{BelZak}Belinskii, V.A., Zakharov, V.E., Integration of the Einstein equations by the methods of inverse scattering theory and construction of explicit multisoliton solutions, Sov. Phys.JETP {\bf 48} 985-994 (1978)
\bibitem{Maison}Maison, D., Are the stationary, axially symmetric Einstein equations completely integrable? Phys. Rev. Lett. {\bf 41},  521--522 (1978)
\bibitem{Bianchi}Bianchi, L., Lezioni di geometria differenziale, Pisa (1909)
\bibitem{Dima}Korotkin, D., On some integrable cases in surface theory, J. Math. Sci. {\bf 94}, 1177--1217 (1999) (Translated from the Russian)
\bibitem{BuMiZa}Burtsev, S. P., Zakharov, V. E., Mikhailov, A. V., The inverse problem method with a variable spectral parameter, Theoret. and Math. Phys. {\bf 70}, 227--240 (1987) 
\bibitem{KokKor}Kokotov, A., Korotkin, D.,
A new hierarchy of integrable systems associated to Hurwitz spaces, math-ph/0112051
\bibitem{Rauch} Rauch, H.E., Weierstrass points, branch points, and moduli of Riemann surfaces, Comm. Pure Appl. Math. {\bf 12}, 543-560 (1959)
\bibitem{Hurtubise} Hurtubise J.C., The Geometry of generalized Hitchin systems, in ``CRM Proceedings and Lecture Notes" {\bf 26}, 55-76 (1999)
\bibitem{Tyurin}Tjurin, A. N., Geometry of moduli of vector bundles, Uspehi Mat. Nauk {\bf 29}, 59--88 (1974)
\bibitem{Fay92} Fay, J., Kernel functions, analytic torsion, and moduli spaces, Memoirs of the American Mathematical Society {\bf 96}, n. 464 (1992)
\bibitem{FKS} Ferapontov, E.V., Korotkin, D.A., Shramchenko V.A., Boyer-Finley equation and systems of hydrodynamic type, Classical Quantum Gravity {\bf 19}, L205-L210 (2002)
\end{thebibliography}
\end{document}